# Radiative enhancement of single quantum emitters in WSe$_2$ monolayers using site-controlled metallic nano-pillars

Tao Cai,[a,b] Je-Hyung Kim,[a] Zhili Yang,[a] Subhojit Dutta,[a] Shahriar Aghaeimeibodi,[a] Edo Waks[a,b]

[a] Department of Electrical and Computer Engineering and Institute for Research in Electronics and Applied Physics, University of Maryland, College Park, Maryland 20742, USA.

[b] Joint Quantum Institute, University of Maryland and the National Institute of Standards and Technology, College Park, Maryland 20742, USA.

* Corresponding author E-mail: edowaks@umd.edu

**ABSTRACT:** Plasmonic nano-structures provide an efficient way to control and enhance the radiative properties of quantum emitters. Coupling these structures to single defects in low-dimensional materials provides a particularly promising material platform to study emitter-plasmon interactions because these emitters are not embedded in a surrounding dielectric. They can therefore approach a near-field plasmonic mode to nanoscale distances, potentially enabling strong light-matter interactions. However, this coupling requires precise alignment of the emitters to the plasmonic mode of the structures, which is particularly difficult to achieve in a site-controlled structure. We present a technique to generate quantum emitters in two-dimensional tungsten diselenide (WSe$_2$) coupled to site-controlled plasmonic nano-pillars. The plasmonic nano-pillar induces strains in the two-dimensional material which generates quantum emitters near the high-field region of the plasmonic mode. The electric field of the nano-pillar mode is nearly parallel to the two-dimensional material and is therefore in the correct orientation to couple to the emitters. We demonstrate both an enhanced spontaneous emission rate and increased brightness of emitters coupled to the nano-pillars. This approach may enable bright site-controlled non-classical light sources for applications in quantum communication and optical quantum computing.

Quantum emitters play an important role in many quantum optical applications such as secured quantum communication,[1,2] quantum computation[3,4] and quantum metrology.[5,6] Single-defect emitters in atomically thin materials[7-16] constitute a new class of non-classical light sources that provides a two-dimensional platform to realize these quantum optical applications. Because they are two-dimensional in nature, these emitters do not suffer from dielectric screening effects and losses due to total internal reflection that hinder other solid-state emitters that are embedded in a material with a high refractive index. Moreover, they can be generated at desired locations by strain-engineering,[17-23] which offers a simple and scalable approach to couple them to photonic devices.

Plasmonic nanostructures constitute an important class of photonic devices for coupling to quantum emitters.[24] These nanostructures concentrate light to sub-wavelength dimensions,[25,26] resulting in strong light-matter interactions that can enhance the brightness and emission rate of quantum emitters[27,28] and achieve strong optical nonlinearities.[29] Furthermore, emitter-plasmon structures offer the possibility to create optical devices that are significantly smaller than the wavelength of light[25,26] and operate at extremely high bandwidths.[30,31] Two-dimensional emitters constitute a particularly promising class of light sources to couple to plasmonic nanostructures because they are not embedded in a dielectric material. They can therefore approach the plasmonic mode of a nanostructure to nanometer distances without being obstructed by the surrounding substrate. But achieving this coupling is challenging due to stringent alignment requirement between the emitter and plasmonic mode which is in the nanometer size scale.[25,26] Such an accurate alignment in a site-localized device is particularly challenging because it requires accurate and simultaneous spatial control of the plasmonic mode and emitter. Atomic force microscope tips can position plasmonic nanostructures close to a quantum emitter,[32] but this approach requires a complicated experimental apparatus and is not easily scalable to multiple devices. Other methods employ an emitter positioned on a dense array of plasmonic nanostructures[33] or generated at a roughened metallic surface[34] in order to ensure the emitter is coupled to at least one localized plasmonic mode, but these approaches are still random and require a large device area. Metal nanowires can also efficiently couple to two-dimensional emitters,[22] but exhibit delocalized propagating surface plasmons with large mode volumes, as opposed to localized plasmons that support tight field confinement and strong light-matter coupling. The ability to create site-localized plasmon-emitter structures with high efficiency remains a difficult problem.

Here we present a technique to generate quantum emitters in two-dimensional semiconductors coupled to site-controlled plasmonic nano-pillars. A lithographically defined plasmonic nano-pillar induces strains in a tungsten diselenide (WSe$_2$) monolayer, leading to the generation of single-defect emitters self-aligned to the plasmonic mode. By taking statistics over many devices, we demonstrate an increased brightness and enhanced spontaneous emission rate in the emitters coupled to plasmonic nano-pillars. These results confirm that plasmonic nano-pillars induce strong light-matter coupling. Our results could find applications in nano-scale ultrafast single photon sources with controlled positions,[27,28] as well as compact nonlinear optical devices operating near the single photon level.[29,35-38]

Figure 1(a) shows a schematic of the plasmonic nano-pillar device design, which is composed of a silicon nano-pillar with a 10 nm- thick gold layer and a 6 nm- thick aluminum oxide (Al$_2$O$_3$) layer on top. We introduce the Al$_2$O$_3$ thin film as a buffer layer to prevent quenching of the emission of WSe$_2$ monolayer,[39] as well as to passivate it to reduce spectral wandering and blinking.[15] The nano-pillar has a diameter of 180 nm and an overall height of about 300 nm. Figure 1(b) shows the numerically calculated electric field distribution near the top of the pillar, determined using a finite-different time-domain (FDTD) method (FDTD solutions, Lumerical). The simulation shows a large field enhancement near the top of the nano-pillar.

Figure 2(a) illustrates the device fabrication procedure. To prepare the sample, we first spin coat negative resist (ma-N 2401, MicroChem) on the silicon substrate and pattern the resist using e-beam lithography. We then transfer the pattern from the resist to the silicon substrate using inductively coupled plasma (ICP) dry etching. We deposit the gold layer using thermal evaporation, followed by atomic layer deposition to deposit the $Al_2O_3$ buffer. The final device consists of a 30x30 array of nanopillars arranged in a square matrix, separated by 4 μm. Figure 2(b) shows a scanning electron micrograph of a small region of the fabricated device, with the inset showing a close-up of a single nano-pillar. Following the pillar fabrication, we transfer $WSe_2$ monolayers synthesized by chemical vapor deposition on a sapphire substrate,[40] onto the pillars using a polydimethylsiloxane (PDMS) substrate as an intermediate transfer medium.[41] Figure 2(c) shows a scanning electron micrograph of one plasmonic nano-pillar covered by a flake of $WSe_2$.

To characterize the sample optically, we first cool it to a base temperature of 3 K using a closed-cycle refrigerator (attoDRY, Attocube Inc.). We perform all photoluminescence measurements using a confocal microscope. An objective lens with a numerical aperture of 0.8 serves to both focus the excitation laser and collect the emitted fluorescence signal. By adjusting the collimation of the input laser, we control the size of the focus to attain either a small diffraction-limited spot that excites a single nano-pillar, or a large spot that excites an area covering multiple pillars for wide-field imaging using a CCD camera (Rolera-XR, Qimaging Inc.). A 715 nm long-pass optical filter (Semrock Inc.) rejects the pump wavelength to isolate the fluorescence signal. For high resolution spectral measurements, we use a grating spectrometer (SP2750, Princeton Instruments). The output of the spectrometer connects to a Hanbury Brown and Twiss intensity interferometer that performs two-photon correlation and lifetime measurements.

Figure 3(a) shows a photoluminescence intensity map of the sample. We excite the sample using a continuous-wave laser emitting at 532 nm, with a large excitation spot that covers multiple nano-pillars. Each nano-pillar introduces a deformation in the atomically thin $WSe_2$ that covers it, leading to the generation of strain-induced single defects in close proximity to the plasmonic mode,[17-23] as evidenced by the bright emission from all nano-pillars in the array. To confirm that the emission originates from single defects, we measure the photoluminescence spectrum from single nano-pillar using a tightly focused spot. Figure 3(b) shows the spectrum of a representative nano-pillar that exhibits a sharp emission line. By fitting the spectrum to a single Lorentzian function, we calculate a linewidth of 0.55 nm for this particular emitter. The spectrum of the observed emitter is distinct from the photoluminescence coming from the bare $WSe_2$ monolayer area which exhibits a broadband emission (see supporting information, Figure S1). Depending on the emitter, we can observe both singlet and doublet spectral emission lines (see supporting information, Figure S2). Both line-shapes are consistent with previously reported spectra from defects in $WSe_2$ monolayer.[8-12] Because emitters are formed by a strain-driven process, the number of emitters in each nano-pillar can vary. Figure 3(b) shows an example where we observe only a single peak corresponding to one emitter, while other pillars exhibit multiple peaks (see supporting information, Figure S3).

In Figure 3(c) we plot the emission intensity of the emitter shown in Figure 3(b) as a function of the excitation power, using a 532 nm continuous-wave excitation source. The intensity shows a saturation behavior that is consistent with emission from a localized single emitter. We fit the measured data to a saturation function of the form $I = I_{sat}P/(P_{sat} + P)$, where $I$ and $I_{sat}$ are the integrated intensity and the saturation intensity, respectively, and $P$ and $P_{sat}$ are excitation power and saturation power, respectively. In the fit, we treat $I_{sat}$ and $P_{sat}$ as fitting parameters. From the fit we determine a saturation power of 0.31 W/cm² (before the objective lens) and a saturation intensity of 8.50×10⁴ counts/s on the single photon counting module.

To further validate that the emission originates from a single defect and show that the defect acts as a quantum light source, we perform a second-order correlation measurement. Figure 3(d) shows the second-order correlation measurement under continuous-wave excitation of the emitter shown in Figure 3(b). The solid line is a numerical fit to a double exponential function. The value of second-order correlation drops below the threshold for photon anti-bunching of 0.5, which confirms that the emission originates from a single photon emitter.

From the exponential fitting to the curve in Figure 3(d), we calculate a lifetime of 0.8 ns which is shorter than the usual lifetimes (a few nano-seconds) of single defects in WSe$_2$ monolayer.[8-12,17-21] The reduced lifetime suggests an enhanced spontaneous emission rate (Purcell enhancement) due to coupling to the surface plasmon mode. However, even in a bare $WSe_2$ monolayer the single-defect emitters exhibit a large variation of lifetime and intensity (see supporting information, Figure S4). Thus, it is difficult to conclude from the measurement of a single emitter whether a Purcell enhancement is present. To do so requires a statistical average of the lifetimes of many emitters. We perform these statistics using measurements from 46 different emitters. We compare the lifetimes of these emitters to a similar number of emitters in two control groups, one in the bare material with no nano-pillars, and the other with nano-pillars that are not coated with gold and therefore exhibit no plasmonic mode confinement. To measure the lifetimes, we excite the emitters using 710 nm laser pulses with a 2 ps- pulsewidth, aligning to the exciton of the $WSe_2$ monolayer (see supporting information, Figure S1), and directly measure the fluorescence decay by a time-resolved lifetime measurement.

Figure 4(a) plots the distribution of lifetimes of the emitters in the plasmonic nano-pillars, where we observe an average lifetime of 2.2±1.5 ns, with a median lifetime of 1.8 ns. As a comparison, Figure 4(b) plots the distribution of lifetimes of 31 emitters in the non-plasmonic nano-pillars (bare silicon nano-pillars), exhibiting an average lifetime of 5.2±2.1 ns with a median lifetime of 4.9 ns. Similarly, the 39 single-defect emitters naturally existing in $WSe_2$ monolayer show an average lifetime of 5.3±2.3 ns with a median lifetime of 5.0 ns, shown in Figure 4(c). The observed shortened lifetimes of emitters in the plasmonic nano-pillars indicate the presence of a Purcell effect, with an average value of 2.4.

We also statistically investigate the photoluminescence emission intensities of the single defects. To measure the emission intensities, we excite the emitters using a continuous-wave laser at 710 nm and measure the emission intensities at saturation using the spectrometer. Figure 5 plots the distribution of emission intensities of three groups of emitters. The emitters in close proximity to the plasmonic structures exhibit a median intensity of $4.4 \times 10^4$ counts/s. The emitters induced by non-plasmonic nano-pillars and the natural emitters exhibit a median intensity of $2.3 \times 10^4$ counts/s and $0.6 \times 10^4$ counts/s respectively. The observed increase in the emission intensities of the plasmonic nano-pillar induced emitters, combined with the increased decay rate, indicate an enhanced radiative decay rate of the emitters due to the coupling to the surface plasmon mode. We note that a number of emitters in the plasmonic nano-pillars exhibit intensities below $5.0 \times 10^4$ counts/s, similar to that of the emitters in the other two control groups. The distribution also exhibits a sharp transition between the range of emitters below and above $5.0 \times 10^4$ counts/s. We attribute this behavior to the fact that some emitters are not created at the high field region of the nano-pillars, and thus show no significant radiative emission enhancement.

In conclusion, we presented a technique to generate quantum emitters in atomically thin semiconductors coupled to site-controlled plasmonic nano-pillars. The lithographically defined plasmonic nano-pillar induced strains in the $WSe_2$ monolayer, which formed single-defect emitters in close proximity to the plasmonic structure. Studies of multiple nano-pillars revealed emitters with statistically shortened lifetimes and increased emission intensities as compared to those not coupled to plasmonic structures, indicating Purcell enhancement. The technique we presented here could be used to realize ultrafast non-classical light sources[27,28] and nonlinear photonic devices.[29] The capability to control the position of the light source enables device integration with more complex quantum optical circuits and offers the possibility for scalable device fabrication. The technique is versatile and could be applied to construct coupled devices composed of various plasmonic nano-structures[24] and single defects in diverse atomically thin materials.[7-16]

## ASSOCIATED CONTENT

**Supporting information**

The following files are available free of charge: Photoluminescence spectrum of a bare $WSe_2$ monolayer area, photoluminescence spectra of representative nano-pillar induced single defects, photoluminescence spectra of representative nano-pillars exhibiting multiple peaks, lifetimes versus photoluminescence intensities of natural single defects in $WSe_2$ monolayer (PDF)

## AUTHOR INFORMATION

**Corresponding Author**

*E-mail: edowaks@umd.edu.

**Funding Sources**

The authors acknowledge support from the National Science Foundation (award number ECCS1508897), the Office of Naval Research ONR (award number N000141410612), the Air Force Office of Scientific Research (AFOSR) (award number 271470871D), and the Physics Frontier Center at the Joint Quantum Institute.

**Notes**

The authors declare no competing financial interest.

FIGURES

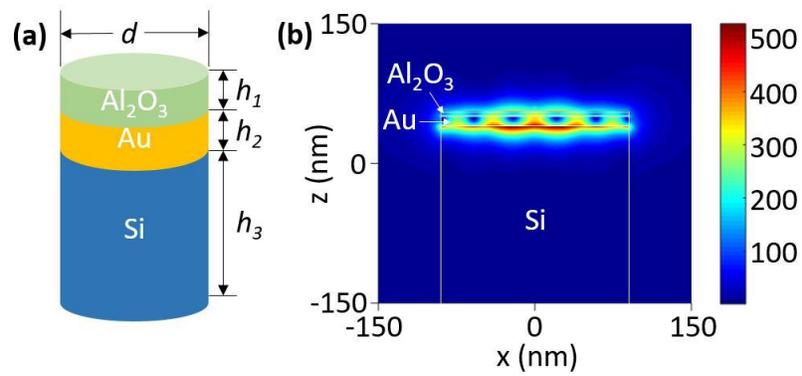

Figure 1. (a) Schematic layout of a single plasmonic nano-pillar. $d$ = 180 nm, $h_1$ = 6 nm, $h_2$ = 10 nm, $h_3$ = 280 nm. (b) Simulated distribution of electric filed *|E|* at the top of the plasmonic nano-pillar.

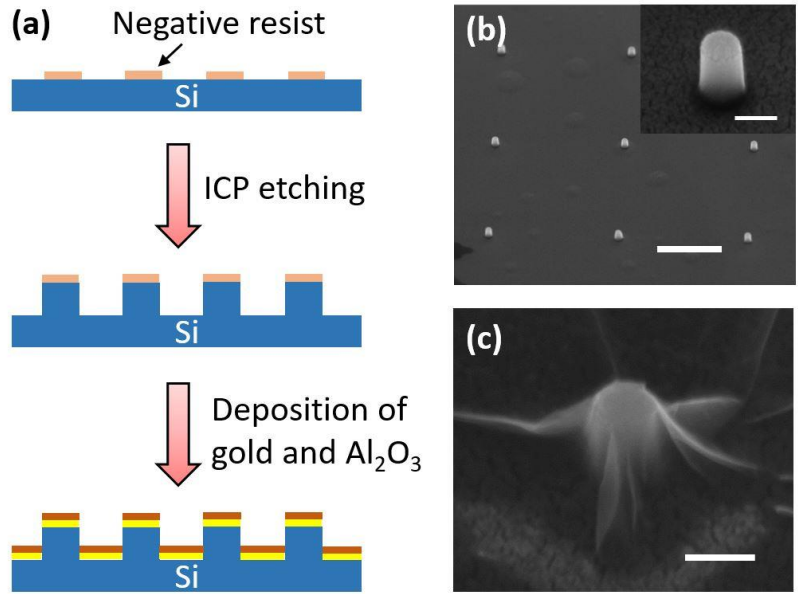

Figure 2. (a) Flow chart of the fabrication procedure of the plasmonic nano-pillars. (b) Scanning electron micrograph showing a part of the array of plasmonic nano-pillars. Scale bar: 2 µm. Inset: close-up of a single plasmonic nano-pillar. Scale bar: 200 nm. (c) Scanning electron micrograph showing a single plasmonic nano-pillar covered by WSe$_2$ monolayer. Scale bar: 200 nm.

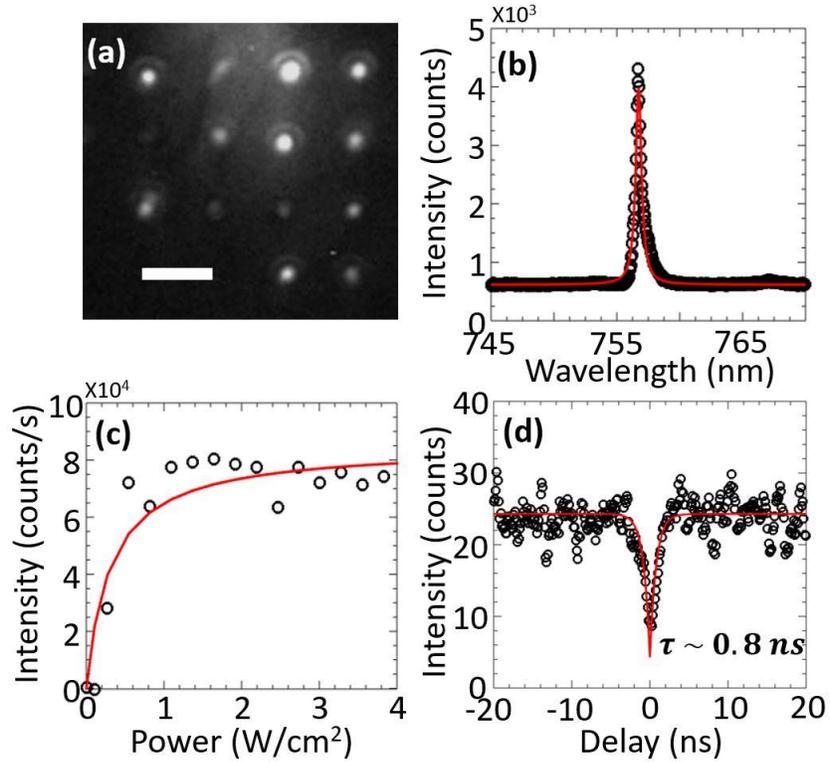

Figure 3. (a) Photoluminescence intensity map of arrays of plasmonic nano-pillars covered by WSe$_2$ monolayer. Scale bar: 4 μm. (b) Photoluminescence spectrum of a representative emitter (black circle) fitted to a single Lorentzian function (solid red curve). (c) Emission intensity of the emitter in panel(b) as a function of the excitation power (black circle), fitted to a saturation function (solid red curve). (d) Second-order correlation measurement of the emitter in panel(b) (black circles) fitted to a double exponential decay function (solid red curve). The fitting exhibits a lifetime of 0.8 ns of this emitter.

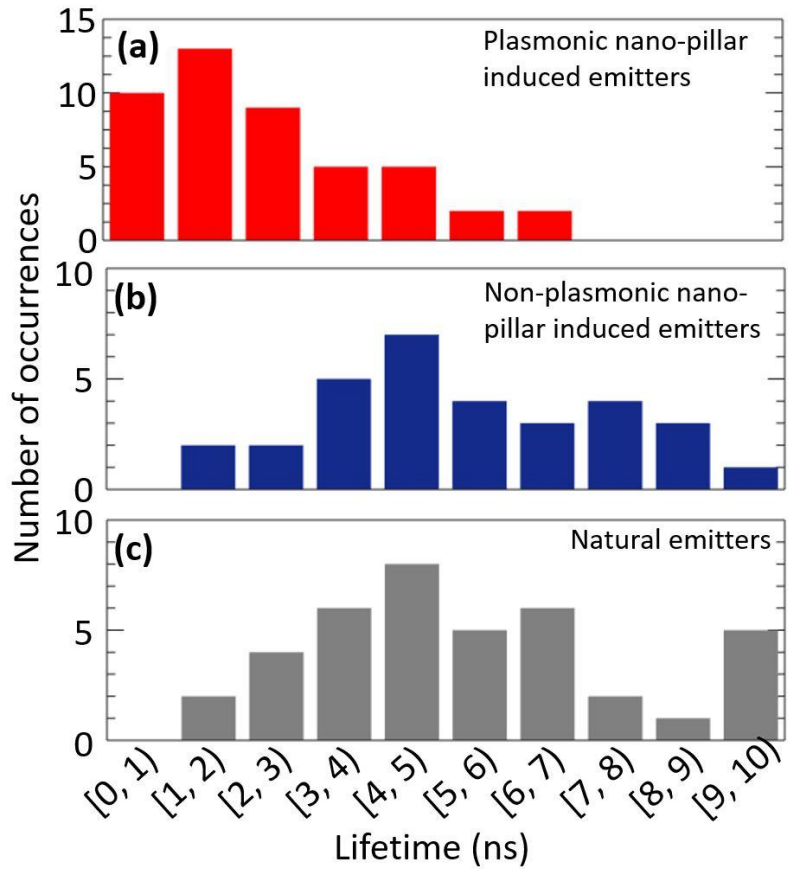

Figure 4. Statistical comparison over lifetimes of single-defect emitters in $WSe_2$ monolayer. Distribution of lifetimes of (a) single defects induced by plasmonic nano-pillars. (b) single defects induced by non-plasmonic nano-pillars. (c) natural single defects.

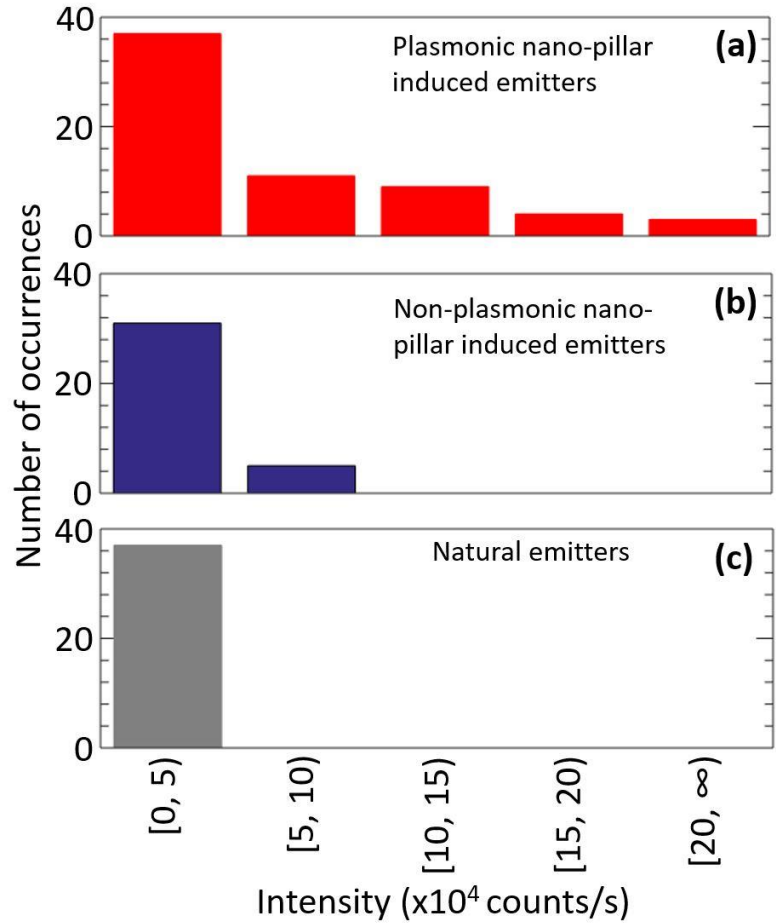

Figure 5. Statistical comparison over photoluminescence emission intensities of single-defect emitters in WSe$_2$ monolayer. Distribution of intensities of (a) single defects induced by plasmonic nano-pillars (b) single defects induced by non-plasmonic nano-pillars. (c) natural single defects.

TABLE OF CONTENTS GRAPH

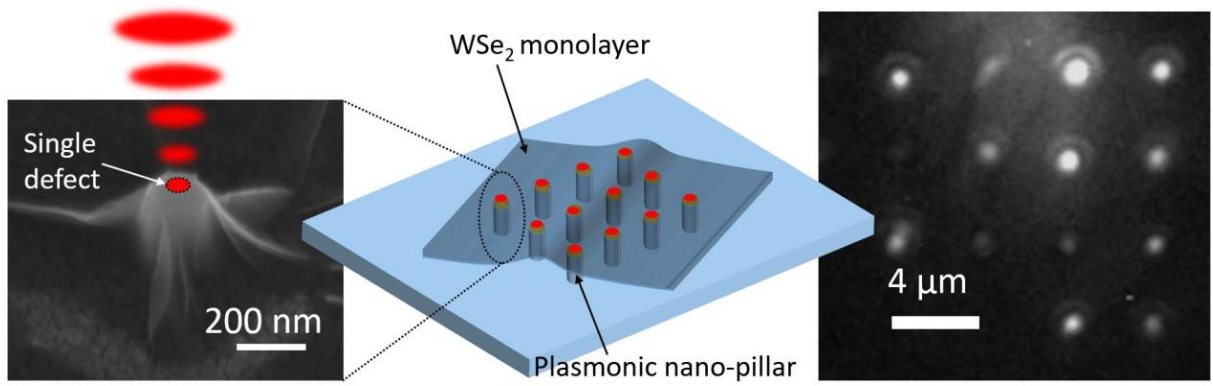